\documentclass[12pt]{article}
\usepackage{graphicx}

\newcommand\pubnumber{}
\newcommand\pubdate{\today}

\def\glasgow{SUPA, Department of Physics and Astronomy\\
University of Glasgow, Glasgow G12 8QQ, UK}

\def\Title#1{\begin{center} {\Large #1 } \end{center}}
\def\Author#1{\begin{center}{ \sc #1} \end{center}}
\def\Address#1{\begin{center}{ \it #1} \end{center}}

\newcommand\pubblock{\rightline{\begin{tabular}{l} \pubnumber\\
         \pubdate  \end{tabular}}}
\newenvironment{Abstract}{\begin{quotation}  }{\end{quotation}}
\newenvironment{Presented}{\begin{quotation} \begin{center} 
             PRESENTED AT\end{center}\bigskip 
      \begin{center}\begin{large}}{\end{large}\end{center} \end{quotation}}

\def\beq{\begin{equation}}
\def\eeq#1{\label{#1}\end{equation}}
\def\eeqn{\end{equation}}

\def\beqa{\begin{eqnarray}}
\def\eeqa#1{\label{#1}\end{eqnarray}}
\def\eeqan{\end{eqnarray}}

\let\bar=\overbar

\def\Dslash{\not{\hbox{\kern-4pt $D$}}}
\def\dslash{\not{\hbox{\kern-2pt $\del$}}}

\def\msb{{\bar{\ssstyle M \kern -1pt S}}}

\begin{document}
\begin{titlepage}
\pubblock

\vfill
\Title{Lattice inputs for the
determination of $|V_{cd}|$ and $|V_{cs}|$
from (semi-)leptonic decays}
\vfill
\Author{Jonna Koponen}
\Address{\glasgow}
\vfill
\begin{Abstract}
This paper is a review of recent lattice QCD results for $D$ and $D_s$ 
meson leptonic and semileptonic decays. The theory inputs needed for
the determination of $V_{cd}$ and $V_{cs}$ from experimental results are 
the meson decay constants (leptonic decays) and the form factors 
(semileptonic decays). In addition one can compare the shape of the form 
factors from lattice QCD and experiment, and use the full experimental 
$q^2$ range (partial decay rates in $q^2$ bins) to determine the CKM matrix 
elements.
\end{Abstract}
\vfill
\begin{Presented}
The 8th International Workshop on the CKM Unitarity Triangle (CKM 2014),
Vienna, Austria, September 8-12, 2014
\end{Presented}
\vfill
\end{titlepage}
\def\thefootnote{\fnsymbol{footnote}}
\setcounter{footnote}{0}

\section{Leptonic and semileptonic decays}

In a leptonic decay a meson (here $D$ or $D_s$) decays to a lepton and its neutrino via a 
virtual $W$ boson. The decay rate is given by
\begin{equation}
\Gamma^{D_s \to \ell\nu} = \frac{G_F^2}{8\pi}m_\ell^2M_{D_s}
\Bigg(1-\frac{m_\ell^2}{M_{D_s}^2}\Bigg)^2f^2_{D_s}|V_{cs}|^2,
\end{equation}
and $(f_{D_s}|V_{cs}|)^2$ can thus be cleanly extracted from experiment. 

On the other hand, consider a semileptonic decay where a $D$ meson decays to a
kaon (or a pion), a lepton and its neutrino. The partial decay rate for a decay
where both the initial and final state mesons are pseudoscalars can be written as
\begin{equation}
\frac{\mathrm{d}\Gamma^{D \to K}}{\mathrm{d}q^2}=\frac{G_F^2p^3}{24\pi^3}
|V_{cs}|^2|f^{D \to K}_+(q^2)|^2.
\end{equation}
Here $p=|\vec{p}|$ is the momentum of the $K$ meson in the rest frame of the $D$,
and $q^2$ is the four-momentum transfer between the two mesons, $q^2=(M_{D}-E_K)^2-p^2$.
In this case experiment can tell us $(|V_{cs}||f^{D \to K}_+(q^2)|)^2$ in a given $q^2$ bin. 

The meson decay constants $f_D$ and $f_{D_s}$ and the form factors $f_+(q^2)$ for the
semileptonic decays can be calculated non-perturbatively in lattice QCD from
first principles. Combining these experimental and theoretical results allows
us to determine the corresponding elements of the CKM matrix, providing
a test of the Standard Model and constraints for new physics. The aim of this
review is to summarize recent lattice QCD results and extract $|V_{cs}|$ and 
$|V_{cd}|$.

\section{Lattice results: decay constants, form factors}

\begin{figure}
\centering
\includegraphics[width=0.99\textwidth]{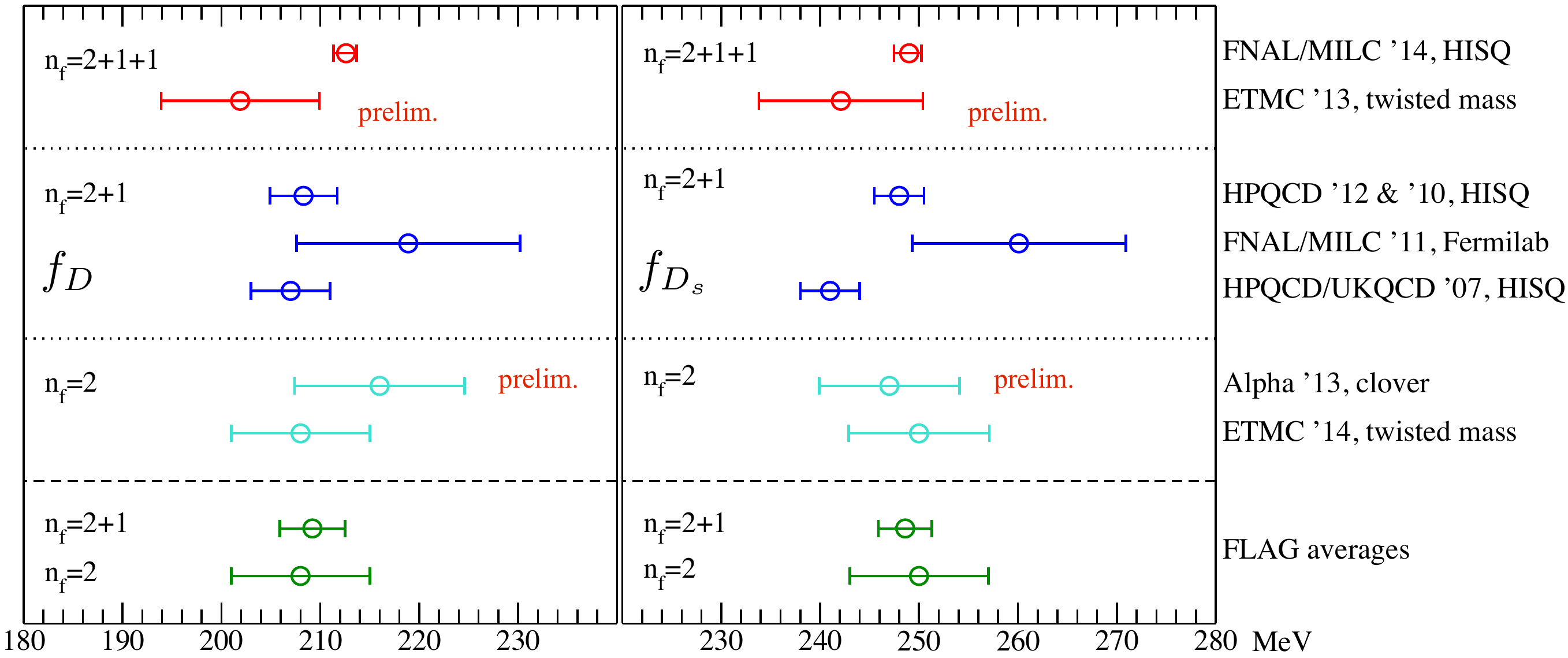}
\caption{Decay constants: $f_D$ on the left, $f_{D_s}$ on the right.
Different discretizations ``HISQ'', ``twisted mass'', ``Fermilab'' and
``clover'' should all agree in the continuum limit, and the agreement
is seen to be good. The best results at the moment, i.e. results with
smallest errors and most modern lattice configurations ($n_f=2+1+1$,
physical pion mass), are from
Fermilab Lattice and MILC Collaborations (FNAL/MILC '14):
$f_D = 212.6 \pm 0.4_{\textrm{stat}} {}^{+1.0}_{-1.2}\big |_{\textrm{syst}}$~MeV and
$f_{D_s} = 249.0 \pm 0.3_{\textrm{stat}} {}^{+1.1}_{-1.5}\big |_{\textrm{syst}}$~MeV.
For completeness, averages from Flavor Lattice Averaging Group (FLAG)~\cite{FLAG} 
are also shown for $2$ and $2+1$ flavors.
}
\label{fig:fDfDs}
\end{figure}

The current status of calculations of the $D$ and $D_s$ meson decay constants
is shown in Fig.~\ref{fig:fDfDs}, tagged by the names of the
lattice groups. The results are 
from~\cite{FNAL14,ETMC13,HPQCD12,HPQCD10,FNAL11,HPQCD07,Alpha13,ETMC14}.
Note that some of the results
are still preliminary. Here $n_f$ denotes the number of flavors used in the calculation:
$n_f=2$ is two light quarks in the sea ($u$ and $d$ quarks that both have the same
mass), $n_f=2+1$ means light and strange quarks and $n_f=2+1+1$ has in addition
charm quarks in the sea. Different discretizations of the Dirac equation for quarks
are denoted by the tags ``HISQ'', ``twisted mass'', ``Fermilab'' and ``clover'' ---
these should all agree in the continuum limit.

\begin{figure}
\centering
\includegraphics[width=0.99\textwidth]{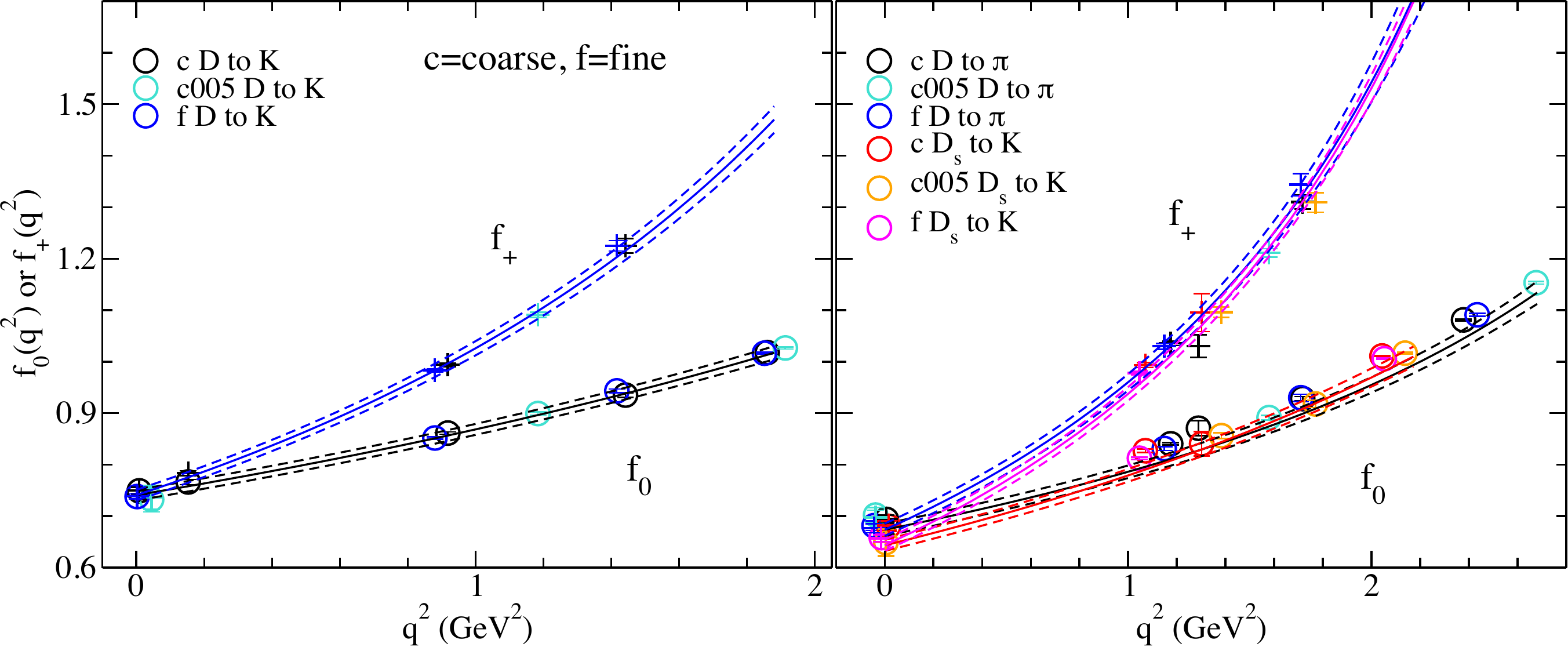}
\caption{On the left: Scalar and vector form factors of $D\to K\ell\nu$ semileptonic
decay~\cite{HPQCD13}. Note the kinematic constraint $f_+(0)=f_0(0)$.
On the right: Form factors of $D\to \pi\ell\nu$ and 
$D_s\to K\ell\nu$ semileptonic decays. Note that the shape of the form factor
is insensitive to the mass of the spectator quark: the form factors for the two
$c\to l$ decays are the same within $\sim 3$ percent. This
has been shown to hold for $B \to D\ell\nu$ and $B_s \to D_s\ell\nu$ 
as well~\cite{Bsemilept}.}
\label{fig:FFs}
\end{figure}

Let us now turn to $D$ meson semileptonic decays:
There are two form factors associated with a pseudoscalar to
pseudoscalar semileptonic decay, a scalar form factor $f_0(q^2)$ and 
a vector form factor $f_+(q^2)$. The scalar form factor is not accessible 
in experiment as it is suppressed in the decay rate by the lepton mass. 
However,
it is quite straightforward to consider a scalar and  a vector current
on the lattice and calculate both form factors.

Here we will only consider lattice results for decays $D \to K\ell\nu$ 
and  $D\to \pi\ell\nu$. Several lattice groups have calculated 
the form factors --- see
Refs.~\cite{HPQCD13,HPQCD11,FNAL04,FNAL_MILC_DK,Sanfilippo}. Fig.~\ref{fig:FFs}
shows results by HPQCD from different lattice spacings [coarse ($a=0.12$~fm)
and fine ($a=0.09$~fm) lattice] and extrapolation to continuum and physical
light quark mass (more details can be found in~\cite{HPQCD13}).
The shapes of the vector form factors agree well with experiment 
--- see for example Fig.~5 in~\cite{HPQCD13} and Fig.~5 in~\cite{FNAL_MILC_DK}.

\section{$|V_{cs}|$ and $|V_{cd}|$}

\begin{figure}
\centering
\includegraphics[width=0.8\textwidth]{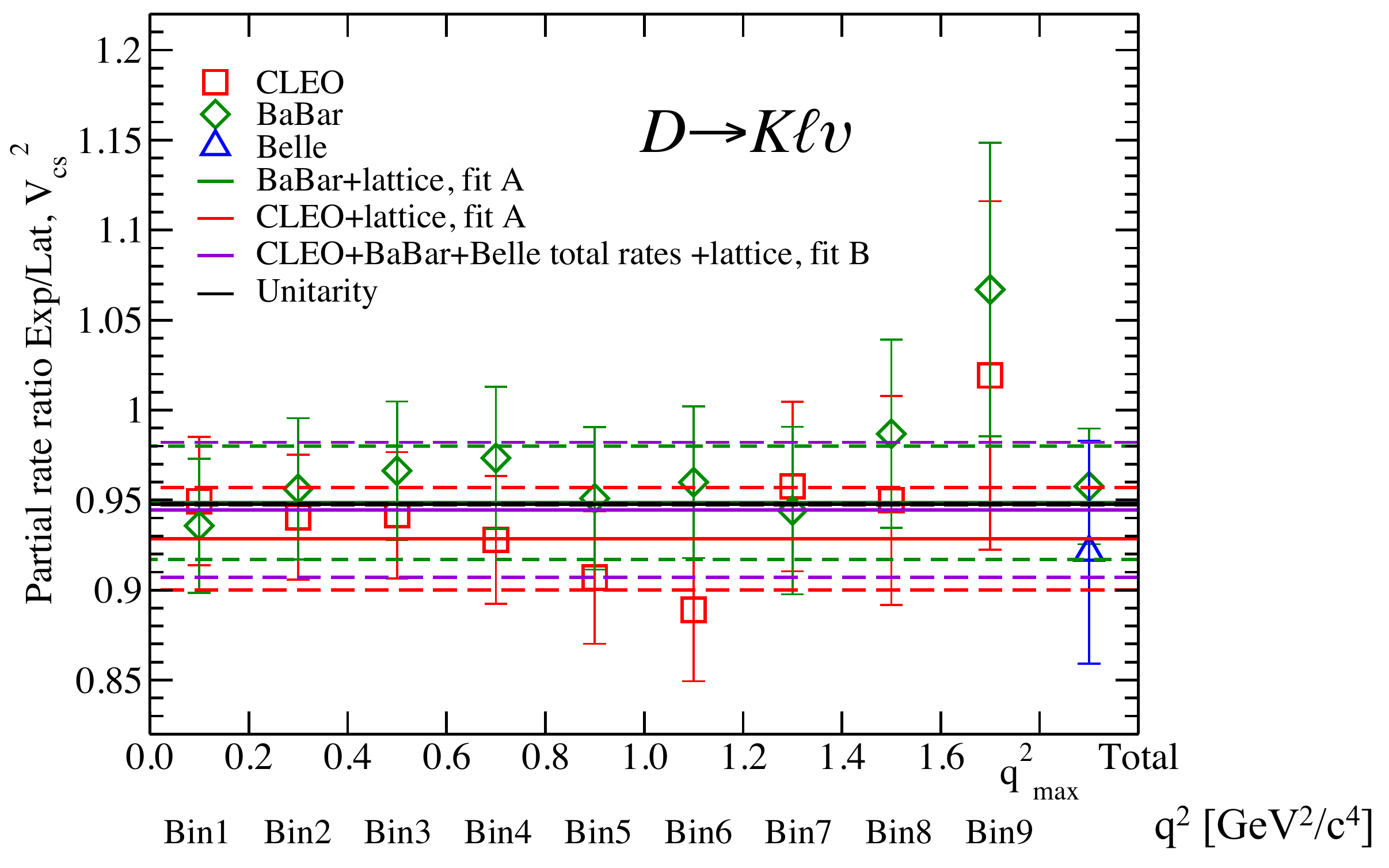}
\caption{$|V_{cs}|$ extracted from $D\to K\ell\nu$ decay using all experimental $q^2$ bins.}
\label{fig:Vcsbin}
\end{figure}

Now we have the needed input from lattice QCD to determine $|V_{cs}|$ and $|V_{cd}|$
from leptonic and semileptonic decays. In the case of a semileptonic decay, we 
can integrate the form factor calculated in lattice QCD over the experimental 
$q^2$ bins and determine the CKM element from each bin: the experimental result
divided by the lattice result for a given bin is $|V_{cs}|^2$ (or $|V_{cd}|^2$)
as shown in Fig.~\ref{fig:Vcsbin}. This also shows that the shape of the form
factor agrees very well between lattice QCD and experiment. One can then
do a weighted average fit to these values, including bin to bin correlations. 
This is more accurate compared to earlier calculations that
extracted CKM elements from experimental knowledge of $(|f_+(0)||V_{cs}|)^2$ 
(or $(|f_+(0)||V_{cd}|)^2$) and a lattice determination of the form factor at $q^2=0$,
since this uses more information.

\begin{figure}
\centering
\includegraphics[width=0.99\textwidth]{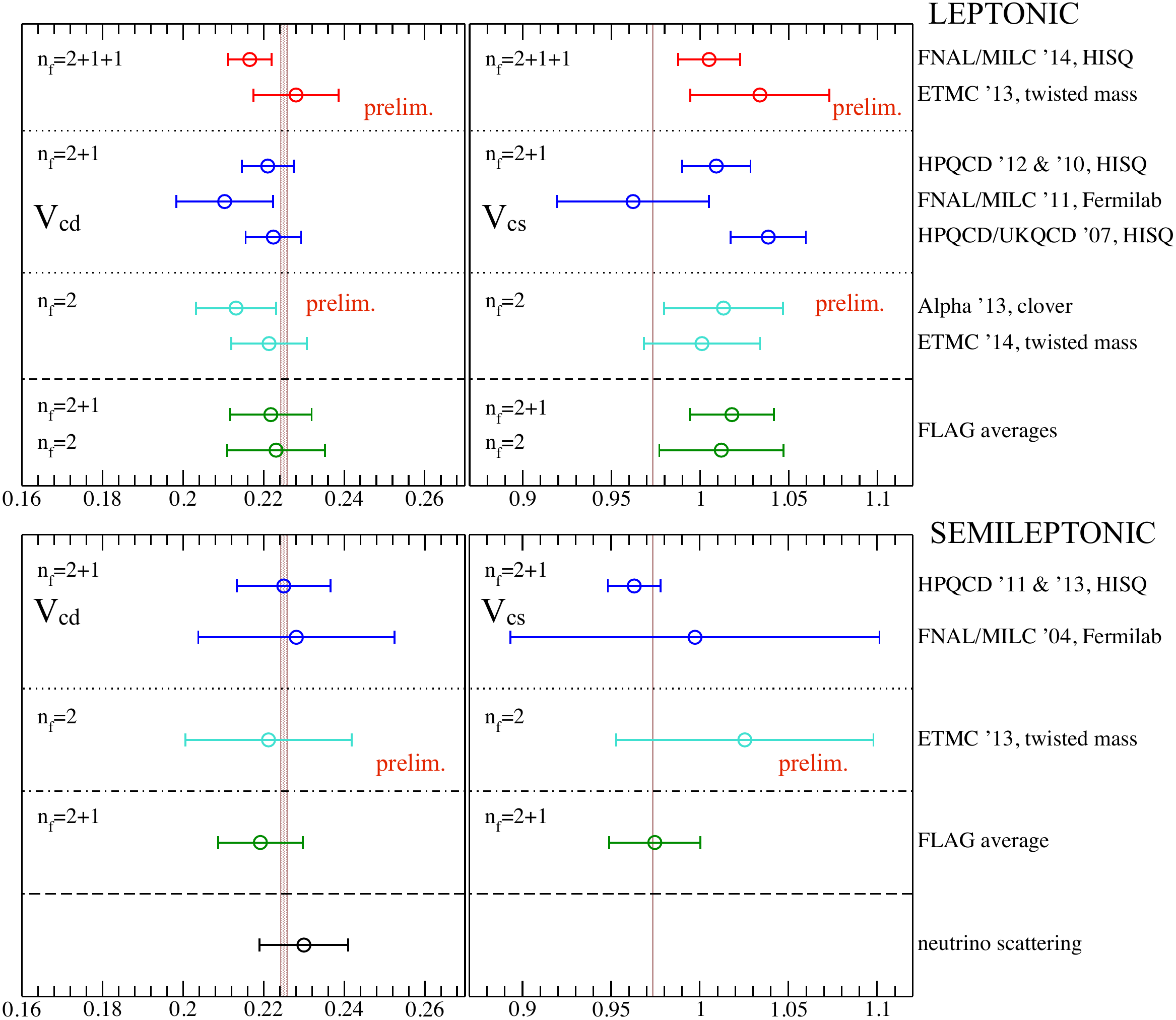}
\caption{Summary of the CKM elements. Top row from left to right:
$|V_{cd}|$ and $|V_{cs}|$ from leptonic decays. Bottom row from left to right:
$|V_{cd}|$ and $|V_{cs}|$ from semileptonic decays. Vertical error
bands show the unitarity value for reference. The best values using
the latest lattice results (most modern lattice configurations with
$n_f=2+1+1$ and physical pion mass, and smallest errors) are: 
$V_{cd}~\textrm{(leptonic)} = 0.2166(52)_{\mathrm{expt}}(13)_{\mathrm{lattice}}$
and $V_{cs}~\textrm{(leptonic)} = 1.005(16)_{\mathrm{expt}}(6)_{\mathrm{lattice}}$, 
taking decay donstants from ~\cite{FNAL14} (FNAL/MILC '14 in Fig.~\ref{fig:fDfDs});
$V_{cd}~\textrm{(semileptonic)} = 0.225(6)_{\mathrm{expt}}(10)_{\mathrm{lattice}}$
from~\cite{HPQCD11} and
$V_{cs}~\textrm{(semileptonic)} = 0.963(5)_{\mathrm{expt}}(14)_{\mathrm{lattice}}$
from~\cite{HPQCD13}. Experimental averages used here are from~\cite{PDG}. 
\cite{HPQCD13} is the first calculation to
use all experimental $q^2$ bins to extract a CKM element from a
semileptonic decay. For comparison, averages from Flavor
Lattice Averaging Group (FLAG)~\cite{FLAG} for $n_f=2$ and $n_f=2+1$
are also shown in the plots, as well as the result for $|V_{cd}|$ from
neutrino scattering experiments~\cite{PDG}.}
\label{fig:VcdVcs}
\end{figure}

The current status of $V_{cd}$ and $V_{cs}$ from leptonic and semileptonic
decays is shown in Fig.~\ref{fig:VcdVcs}. The tags are the same as for the
decay constants. Leptonic decays tend to give a higher value for
$V_{cs}$ than the unitarity value, but note that all data points in
the plot would shift to left or right, if the experimental average
changed. All lattice results agree with each other very well, and the
semileptonic determination of $V_{cs}$ and both leptonic and
semileptonic determinations of $V_{cd}$ agree with the assumption of
CKM matrix unitarity.

\end{document}